\documentclass[english,12pt]{article}
\usepackage[cp1251]{inputenc}
\usepackage{babel}
\usepackage{amsfonts, amsmath}
\usepackage{texnames}
\usepackage[T1]{fontenc}
\usepackage{graphicx}
\newcommand{\be}{\begin{equation}} \newcommand{\ee}{\end{equation}}

\def\A{\ensuremath{\boldsymbol{A}}}

\newcommand{\gsim}{\mathrel{\hbox{\rlap{\lower.55ex\hbox{$\sim$}} \kern-.3em \raise.4ex \hbox{$>$}}}}

\newcommand{\lsim}{\mathrel{\hbox{\rlap{\lower.55ex\hbox{$\sim$}} \kern-.3em \raise.4ex \hbox{$<$}}}}

\begin{document}

\begin{center}
{\bf The  Quantum  Primordial Black Holes,Dimensionless Small Parameter,Inflationary Cosmology and Non-Gaussianity}\\
 \vspace{5mm} Alexander
Shalyt-Margolin \footnote{E-mail: a.shalyt@mail.ru;
alexm@hep.by}\\ \vspace{5mm} \textit{Research Institute for
Nuclear Problems,Belarusian State University, 11 Bobruiskaya str.,
Minsk 220040, Belarus}
\\{\bf  This paper is devoted to the 80-th Anniversary of Professor Vasili Ivanovich Strazhev}
\end{center}

\begin{abstract}
In  the present work consideration is given to the primordial
black holes ({\bf pbhs}) in the Schwarzschild-de Sitter Metric
with small mass (ultralight) in the preinflationary epoch. Within
the scope of natural assumptions, it has been shown that the
quantum-gravitational corrections ({\bf qgcs}) to the
characteristics of such black holes can contribute to all the
cosmological parameters, shifting  them compared with the
semiclassical consideration. These contributions are determined by
a series expansion in terms of a small parameter dependent on  the
hole mass (radius). For this pattern different cases have been
considered (stationary, black hole evaporation...). It has been
demonstrated that involvement of  ({\bf qgcs}) leads to a higher
probability for the occurrence of  such {\bf pbhs}. Besides,
high-energy deformations of Friedmann Equations created on the
basis of these corrections have been derived for different
patterns. In the last section of this work it is introduced a
study into the contributions generated by the above-mentioned {\bf
qgcs} in inflationary  cosmological perturbations. Besides, it has
been shown that non-Gaussianity of these perturbations is higher
as compared to the semi-classical pattern.
\end{abstract}

PACS: 11.10.-z,11.15.Ha,12.38.Bx
\\
\noindent Key words:primordial black holes; inflationary
cosmology; quantum-gravitational corrections;non-gaussianity
 \rm\normalsize \vspace{0.5cm}

\section{Introduction}

The primordial black holes ({\bf pbhs})
\cite{zeldovich-novikov-I}--\cite{carr-hawking} in the Early
Universe are due to gravitational collapse of the high-density
matter \cite{PBH--1}. In \cite{carr-1}--\cite{carr-3} a
sufficiently accurate estimate of the mass {\bf pbhs} $M(t)$
formed during the period of time $t$ since the Big Bang has been
obtained
\begin{equation}\label{PBH-Mass}
M(t)\approx {c^3 t\over G} \approx 10^{15}\left({t\over 10^{- 23}
\ {\rm s}}\right) g.
\end{equation}
As seen, for small times close to the Planckian time $t=t_p\approx
10^{-43}$s, the mass of {\bf pbhs} is close to the Planck mass
$M(t)\approx 10^{-5}g$. The names of such black holes were varying
with time: ''mini-black holes'',''micro-black holes'', and, e.g.
in \cite{Prof-1}, they were referred to as ''ultralight primordial
black holes''. The author of this paper uses for such {\bf pbhs}
the name {\bf quantum pbhs} (or {\bf qpbhs}) introduced in
\cite{Calmet--1},\cite{Calmet--2} and notes that
quantum-gravitational effects for these objects could be
significant. Of particular interest are {\bf pbhs} arising in the
preinflationary epoch. In \cite{Prok} a semiclassical
approximation was used to study the problem of scalar
perturbations due to such {\bf pbhs}. But, considering that all
the processes in this case proceed at very high energies $E$ close
to the Planckian $E\simeq E_p$, the inclusion of quantum-gravity
corrections {\bf qgcs} for these black holes in this pattern is
necessary if quantum gravity exists \cite{Keifer--book}. Despite
the fact that presently there is no self-consistent theory of
quantum gravity, a consensus is reached on correctness of some
approaches to the theory, specifically, replacement of the
Heisenberg Uncertainty Principle {\bf(HUP)} by the Generalized
Uncertainty Principle {\bf(GUP)} on going to high (Planck’s)
energies, used in this paper.
\\Within the scope of a natural assumption based on  the results from \cite{Prok},
in the present work the author  studies the problem, how the
above-mentioned {\bf qgcs} for quantum {\bf pbhs} can shift the
inflationary parameters and contribute to cosmological
perturbations involved in the inflationary process. Moreover, it
is shown that inclusion of {\bf qgcs}:
\\{\bf a}) increases the probability of the generation of  {\bf pbhs};
\\{\bf b}) leads to the enhancement of  non-Gaussianity in cosmological perturbations.
\\ In what follows the abbreviation {\bf qgcs} is associated with the foregoing quantum-gravity  corrections.
Beginning a study of {\bf qgcs} in \cite{Shalyt-Proc}, here the author, using the definitions  from   \cite{Shalyt-Proc},
presents much more general results referring to the emergence probability of quantum {\bf pbhs} as well as of
quantum fluctuations,perturbations and non-Gaussianity.
\\In the last twenty years numerous works devoted to a a primordial  black  holes
mechanisms of their formation and effects.  It is worth noting
both interesting reviews on this topic \cite{PBH-1} and important
results concerning specific problems \cite{PBH-2}-\cite{PBH-4}.
These articles and other works in the field of primordial  black
holes  motivated the present research.
\\The present paper is structured as follows.
\\ Section 2 presents the instruments used to obtain the principal results.
Section 3 shows  how  {\bf qgcs} shift the inflationary parameters
within a natural assumption from \cite{Prok}. In Section 4 it is
demonstrated that inclusion of {\bf qgcs} increases the occurrence
probability for such {\bf pbhs}. In Section 5 the high-energy
deformations of Friedmann Equations on the basis of  {\bf qgcs}
are derived for different cases. Finally, Section 6 begins a study
of the contributions made by {\bf qgcs} into different
cosmological perturbations under inflation and, due to the
involvement of  {\bf qgcs}, demonstrates the enhancement of
non-Gaussianity for different perturbations revealed as growing
moduli of {\bf bispectrums}.
\\In what follows the normalization $c=\hbar=1$ is used, for which we have $G=l^2_p$.

\section{PBHs with the Schwarzschild-de Sitter Metric in the Early Universe}

It should be noted that Schwarzschild black holes  in real physics
(cosmology, astrophysics) are idealized objects. As noted in
(p.324,\cite{Frol}): ''Spherically symmetric accretion onto a
Schwarzschild black hole is probably only of academic interest as
a testing for theoretical ideas. It is of little relevance for
interpretations of the observations data. More realistic is the
situation where a black hole moves with respect to the
interstellar gas...''
\\Nevertheless, black holes just of this type may arise and may be realistic in the early Universe.
In this case they are  {\bf pbhs}.
\\During studies of the early Universe for  such {\bf pbhs}
the Schwarzschild metric \cite{Wald},\cite{Frol}
\begin{equation}\label{BH-0}
ds^{2}=\left( 1-\frac{2MG}{r}\right) dt^{2}-\left(
1-\frac{2MG}{r}\right) ^{-1}dr^{2}-r^{2}d\Omega ^{2},
\end{equation}
for {\bf pbhs} is replaced by the Schwarzschild-de~Sitter (SdS)
metric \cite{Prok} that is associated with Schwarzschild black
holes with small mass $M$ in the early Universe, in particular in
pre-inflation epoch
\begin{equation}\label{SdS-1}
ds^2 = -f(\tilde{r})dt^2 + \frac{d\tilde{r}^2}{f(\tilde{r})} +
\tilde{r}^2d\Omega^2 \,
\end{equation}
where $f(\tilde{r})=1-2GM/\tilde{r}-\Lambda
\tilde{r}^2/3=1-2GM/\tilde{r}-
\tilde{r}^2/L^{2},L=\sqrt{3/\Lambda}=H_0^{-1}$, $M$ - black hole
mass, $\Lambda$ -- cosmological constant, and $L=H_0^{-1}$ is the
Hubble radius.
\\In general, such a black hole may have two different horizons
corresponding to two different zeros $f(\tilde{r})$: event horizon
of a black hole and cosmological horizon. This is just so in the
case under study when a value of $M$ is small
\cite{Nar-1},\cite{Nar-2}. In the general case of $L\gg GM$, for
the event horizon radius of a black hole having the metric
(\ref{SdS-1}), $r_{\rm H}$ takes the following form (formula (9)
in \cite{Cas}):
\begin{equation}\label{r-hor-asym}
 r_{\rm H} \simeq 2GM \left[1 +
\left(\frac{r_M}{L}\right)^2 \right], where\;r_M=2MG.
\end{equation}
Then, due to the assumption concerning the initial smallness of
$\Lambda$, we have $L\gg r_M$. In this case, to a high accuracy,
the condition $r_{\rm H}=r_M$ is fulfilled, i.e. for the
considered (SdS) BH we can use the formulae for a Schwarzschild
BH, to a great accuracy.
\\{\bf Remark  2.1.}
\\{\it Note that, because $\Lambda$ is very small, the condition $L\gg GM$
and hence the formula of (\ref{r-hor-asym}) are obviously valid
not only for black hole with the mass $M\propto m_p$ but also for
a much greater range of masses, i.e. for black holes with the mass
$M\gg m_p$, taking into account the condition $L\gg GM$. In fact
we obtain ordinary Schwarzschild black holes (\ref{BH-0}) with
small radius}.
\\Specifically, for the energies on the order of Plank energies
(quantum gravity scales) $E\simeq E_p$, the Heisenberg Uncertainty
Principle ({\bf HUP}) \cite{Heis1}
\begin{equation}
\left( \delta X\right) \left( \delta P\right) \geq
\frac{\hbar}{2}, \label{Heis}
\end{equation}
 may be replaced by the Generalized Uncertainty Principle ({\bf GUP})
\cite{Nou}
\begin{equation}\label{GUP}
[X,P]=i\hbar\exp \left(\frac{\alpha ^{2}l_{p}^{2}}{\hbar
^{2}}P^{2}\right),\quad \left( \delta X\right) \left( \delta
P\right) \geq \frac{\hbar }{2}\left\langle \exp \left(
\frac{\alpha ^{2}l_{p}^{2}}{\hbar ^{2}}P^{2}\right) \right\rangle.
\end{equation}
Then for saturate ({\bf GUP*}) $\left( \delta
P\right)^{2}=\left\langle P^{2}\right\rangle -\left\langle
P\right\rangle ^{2}$ and in virtue of formula (8) in \cite{Nou}
\begin{equation}\label{GUP.2}
\left( \delta X\right) \left( \delta P\right) =\frac{\hbar
}{2}\exp \left( \frac{\alpha ^{2}L_{P\text{l}}^{2}}{\hbar
^{2}}\left( \left( \delta P\right) ^{2}+\left\langle
P\right\rangle ^{2}\right) \right) .
\end{equation}
\\It should be noted that the validity of {\bf GUP} at Planck’s scales is partly based on the Gedanken Experiment in
Quantum Gravity from quantum (or same micro) black holes
\cite{Scard-1}. In this case, due to {\bf GUP}, the physics
becomes nonlocal and the position of any point is determined
accurate to   $l_{min}$. It is impossible to ignore this
nonlocality at the energies close to the Planck energy $E\approx
E_p$, i.e. at the scales $l\propto l_p$ (equivalently we have
$l\propto r_{min}=l_{min}$).
\\It is further assumed that the  holds true {\bf GUP*} i.e.
(\ref{GUP.2}).
\\Then there is a possibility for existence of Planck Schwarzschild black
hole, and accordingly of a Schwarzschild sphere (further referred
to as ''minimal'') with the minimal mass $M_{0}$ and the minimal
radius $r_{min}$ (formula (20) in \cite{Nou}) that is a
theoretical minimal length $r_{min}$:
\begin{equation}\label{min}
r_{min}=l_{min}=\left( \delta X\right)
_{0}=\sqrt{\frac{e}{2}}\alpha l_{p},\quad M_{0}=\frac{\alpha
\sqrt{e}}{2\sqrt{2}}m_{p},
\end{equation}
where $\alpha$ - model-dependent parameters on the order of 1, $e$
- base of natural logarithms, and  $r_{min}\propto
l_{p},M_{0}\propto m_{p}$.
\\Actually, \cite{Nou} presents calculated values
of the mass $M$ and the radius $R$ for Schwarzschild  BH with
regard to the quantum-gravitational corrections within the scope
of {\bf GUP*} (\ref{GUP.2}).
\\With the use of the normalization $G=l^2_{p}$ adopted in \cite{Nou},
temperature of a Schwarzschild black hole having the mass $M$ (the
radius $R$) \cite{Frol} in a semi-classical approximation takes
the form
\begin{equation}\label{semicl}
T_{M}=\frac{1}{8\pi GM}.
\end{equation}
Within the scope of {\bf GUP*} (\ref{GUP.2}),the temperature
$T_{M}$ taking into account to ({\bf qgc}) is of the form ((23)
and (25) in \cite{Nou}))
\begin{eqnarray}\label{quant-grav}
T_{M,q}=\frac{1}{8\pi MG}\exp\left(-\frac{1}{2}W\left( -%
\frac{1}{e}\left( \frac{M_{0}}{M}\right)
^{2}\right)\right)=\frac{1}{8\pi MG}\exp\left(-\frac{1}{2}W\left( -%
\frac{1}{e}\left( \frac{A_{0}}{A}\right)
\right)\right)=\nonumber\\=\frac{1}{8\pi MG}\left(
1+\frac{1}{2e}\left( \frac{M_{0}}{M}\right)
^{2}+\frac{5}{8e^{2}}\left( \frac{M_{0}}{M}\right)
^{4}+\frac{49}{48e^{3}}\left( \frac{M_{0}}{M}\right) ^{6}+\ldots
\right),
\end{eqnarray}
where $A$ is the black hole horizon area of the given black hole
with mass $M$ and event horizon $r_M$ ,$A_{0}=4\pi\left(\delta
X\right)_{0}^{2}$ is the black hole horizon area of a minimal
quantum  black hole from formula (\ref{min}) and  $W\left( -
\frac{1}{e}\left(\frac{M_{0}}{M}\right)^{2}\right)=W\left( -%
\frac{1}{e}\left( \frac{A_{0}}{A}\right) \right)$ -- value at the
corresponding point of the Lambert W-function $W(u)$ satisfying
the equation (formulae (1.5) in \cite{Lamb} and (9) in \cite{Nou})
\begin{equation}\label{W-new}
W\left( u\right)e^{W\left(u\right)}=u.
\end{equation}
$W\left(u\right)$ is the multifunction for complex variable
$u=x+yi$. However, for real $u=x,-1/e\leq u<0$,$W\left( u\right)$
is the single-valued continuous function having two  branches
denoted by $W_{0}(u)$ and $W_{-1}(u)$ , and  for real $u=x,u\geq
0$ there is only one branch  $W_{0}(u)$ \cite{Lamb}.
\\Obviously, the quantum-gravitational correction
({\bf qgc}) (\ref{quant-grav}) presents a {\it deformation} (or
more exactly, the {\it quantum deformation} of a classical
black-holes theory from the viewpoint of the paper \cite{Fadd}
with the deformation parameter $A_{0}/A$):
\begin{equation}\label{QGC-1.Sch}
\frac{A_{0}}{A}=\frac{4\pi r_{h}^{2}}{4\pi r^{2}_M}=\frac{
l^{2}_{min}}{r^{2}_M},
\end{equation}
where $r_{h}=l_{min}$ is the horizon radius of minimal {\bf pbh}
from formula  (\ref{min}).
\\It should be noted that this deformation parameter
\begin{equation}\label{QGC-1.Sch-old}
l^{2}_{min}/r^{2}_M\doteq \alpha_{r_M}
\end{equation}
has been introduced by the author in his earlier works
\cite{shalyt2},\cite{shalyt3}, where he studied deformation of
quantum mechanics at Planck scales in terms of the deformed
quantum  mechanical density matrix. In the Schwarzschild black
hole case $\alpha_{r_M}=l^{2}_{min}\mathcal{K}$ -- Gaussian
curvature $\mathcal{K}=1/\alpha_{r_M}$ of the black-hole event
horizon surface \cite{Diff-Geom}.
\\It is clear that, for a great black hole having large mass
$M$ and great event horizon area $A$, the deformation parameter
$\frac{1}{e}\left(\frac{M_{0}}{M}\right)^{2}$
is vanishingly small and close to zero. Then a value of $W\left( -%
\frac{1}{e}\left( \frac{M_{0}}{M}\right) ^{2}\right)$ Is also
close to $W(0)$. As seen, $W(0)=0$ is an obvious solution for the
equation (\ref{W-new}). We have
\begin{equation}\label{QGC-1.1}
\exp\left(-\frac{1}{2}W\left(-\frac{1}{e}\left(
\frac{M_{0}}{M}\right)^{2}\right) \right)\approx 1.
\end{equation}
So, a black hole with great mass $M\gg m_p$ necessitates no
consideration of {\bf qgcs}.
\\But in the case of small black holes we have
\begin{equation}\label{QGC-1.1s}
\exp\left(-\frac{1}{2}W\left(-\frac{1}{e}\left(
\frac{M_{0}}{M}\right)^{2}\right) \right)> 1.
\end{equation}
In formulae above it is assumed that $M>M_{0}$, i.e. the black
hole under study is not minimal (\ref{min}).
\\We can rewrite the formula of (\ref{quant-grav}) as follows:
\begin{eqnarray}\label{quant-grav.new}
T_{M_q}=\frac{1}{8\pi M_q G},M_q=M\exp\left(\frac{1}{2}W\left( -%
\frac{1}{e}\left( \frac{M_{0}}{M}\right) ^{2}\right)\right);\nonumber\\
R_q=2M_qG=R\exp\left(\frac{1}{2}W\left( -%
\frac{1}{e}\left( \frac{M_{0}}{M}\right) ^{2}\right)\right),
\end{eqnarray}
where $M_q$ and $R_q$ are respectively the initial black-hole mass
and event horizon radius considering {\bf qgcs} caused by {\bf
GUP*} (\ref{GUP.2}),i.e. $T_{M_q}$ is the temperature of black
hole with mass  $M_q$.
\\Note that in a similar way   $M_q$ is involved in
(\cite{Nou} formula (26)) as a function of the black-hole
temperature. But, instead of the small parameter  $M_{0}/M$, the
author uses the small parameter $T_{H}/T^{max}_{H}$, where
$T^{max}_{H}$ is the black hole maximum temperature.
\\{\bf Remark 2.2}
\\{\it It is clear that the formula (\ref{quant-grav.new})
with the substitution of $M\mapsto M_q$ is of the same form as
formula (\ref{semicl}), in fact representing
(\ref{quant-grav}),i.e. in the formula for temperature of a black
hole  the inclusion of {\bf qgcs} may be realized in two ways with
the same result: (a)the initial mass $M$ remains unaltered and
{\bf qgcs} are involved only in the formula for temperature, in
this case  (\ref{quant-grav}); (b){\bf qgcs} are involved in the
mass-the above-mentioned substitution takes place $M\mapsto M_q$
(formula(\ref{quant-grav.new})). Such ''duality'' is absolutely
right in this case if a black hole is considered in the stationary
state in the absence of accretion and radiation processes. Just
this case is also studied in the paper.
\\A recent preprint \cite{Quant-Inf-1}
in the case  (b) for the space-time dimension  $D\geq 4$, using
approaches to quantum gravity of the alternative  {\bf GUP} ({\bf
GUP*}), gives a formula for the mass $M_q$ of a black hole with a
due regard to {\bf qgc}
\begin{equation}\label{mass}
M_q = \left[1 - \eta \exp\left(-\frac{\pi
r_{0}^{D-2}}{G_{D}}\right)\right]^{D-3}M.
\end{equation}
Here in terms of \cite{Quant-Inf-1} $r_{0}$ is the Schwarzschild
radius of the primordial black hole with the mass
$M$,$G_{D}$-gravitational constant in the dimension $D$, and $\eta
= [0, 1]$ is a parameter. In case under study this parameter, as
distinct from cosmology, has no relation to conformal time.
Obviously, for $\eta=0$ we have a semiclassical approximation and,
as noted in \cite{Quant-Inf-1}, the case when $\eta=1$ corresponds
to {\bf qgc} as predicted by a string theory}.
\\{\bf Remark 2.3}
\\{\it It should be noted that, with
the expansion  $\exp\left(-\frac{1}{2}W\left( -%
\frac{1}{e}\left( \frac{M_{0}}{M}\right) ^{2}\right)\right)$ in
terms of the small parameter  $\alpha_{r_{M}}=(M_{0}/M)^2$, in
formula (\ref{quant-grav}) one can easily obtain the
small-parameter expansion
of the inverse number $\exp\left(\frac{1}{2}W\left( -%
\frac{1}{e}\left( \frac{M_{0}}{M}\right) ^{2}\right)\right)$,
and also of all the integer powers for this exponent,  specifically for its square
 $\exp\left(-W\left( -%
\frac{1}{e}\left( \frac{M_{0}}{M}\right) ^{2}\right)\right)$}.

\section{Inflation Parameters  Shifts  Generated by  QGCs}

To this end in cosmology, in particular inflationary, the metric
(\ref{SdS-1}) is conveniently described in terms of the conformal
time $\eta$ \cite{Prok}:
\begin{equation}\label{SdS-1.conf}
ds^2 = a^2(\eta) \Bigg\{-d\eta^2
     + \left(1+\frac{\mu^3\eta^3}{r^3}\right)^{4/3}
           \left[\left(\frac{1-\mu^3\eta^3/r^3}
                            {1+\mu^3\eta^3/r^3}\right)^2dr^2
     + r^2d\Omega^2\right]\Bigg\},
\end{equation}
where $\mu = (GMH_0/2)^{1/3}$,  $H_0$ -- de Sitter-Hubble
parameter and scale factor, $a$ -- conformal time function $\eta$:
\begin{equation}\label{SdS-1.a}
a(\eta)=-1/(H_0\eta),\eta<0.
\end{equation}
Here $r$ satisfies the condition $r_0<r<\infty$ and a value of
$r_0=-\mu\eta$ in the reference frame of (\ref{SdS-1.conf})
conforms to singularity of the back hole.
\\Due to (\ref{r-hor-asym}), $\mu$ may be given as
\begin{equation}\label{SdS-1.mu}
\mu=(r_{M}H_0/4)^{1/3},
\end{equation}
where $r_{M}$ is the radius of a black hole with the SdS
Schwarzschild-de Sitter metric (\ref{SdS-1}).
\\
\\{\bf  Remark 3.1.}
\\{\it Note that there are no limitations for primordial black holes
in \cite{Prok}, excepting the fact that these are {\bf pbhs} with
the SdS-metric (\ref{SdS-1}) in pre-inflationary era having small
radii (which are actually close to Planck’s values). Similarly to
\cite{Nou}, {\bf qgcs}have been calculated for all the
Schwarzschild black holes within the scope of {\bf GUP*} ({\bf
GUP}), (\ref{GUP.2}),(\ref{GUP}). These formulae are in the most
general form because the principle corresponds to a ''minimal''
deformation (imposing no additional conditions) of the canonical
quantum theory on going to the Planck energies
\cite{Ven3}--\cite{Tawf}. As seen in formula (\ref{QGC-1.1}), for
black holes with great masses, the above-mentioned {\bf qgcs} are
small and may be neglected. Considering formula
(\ref{QGC-1.1s}),they are significant only for {\bf pbhs} having
small radii and this is the case under study}.
\\ As we consider  quantum {\bf pbhs} with the SdS-metric, in
this case, as noted in  (\ref{r-hor-asym}), they to a high
accuracy are  coincident with the Schwarzschild black holes
(\ref{BH-0})  and hence they have the identical formulae for {\bf
qgcs}:(\ref{quant-grav.new}),(\ref{mass})... Further we consider
the contribution made by these {\bf qgcs} into the quantities
associated with inflation: inflationary parameters, cosmological
perturbations etc. It should be made clear:
\\In \cite{Prok} in general only the case $\mu=const$ is considered and,
as noted in \cite{Prok},  for the case $\mu\neq const$ we can use
only the pattern including the radiation processes of {\bf pbhs}.
However,  the value of  $\mu=const$ itself contains no information
concerning the treatment of either the initial quantum {\bf pbh}
in a semiclassical approximation or the consideration with regard
to {\bf qgcs}. Obviously, in this case involvement of {\bf qgcs}
shifts all the parameters derived for a  semiclassical (canonical)
pattern.
\\Let us consider the following pattern related to that studied in \cite{Prok}: it is supposed
that, as the mass $M$ of {\bf pbh} may be changed due to
 the radiation process, the corresponding change takes place for $\mu$ -- in the general case we have ($\mu\neq const$) in view
 of these processes.
 But further it is assumed that after termination of these processes  $\mu$ is unaltered with regard to {\bf
qgcs}, i.e. in formula (\ref{SdS-1.mu}) we have
$\mu=(r_{M}H_0/4)^{1/3}=(r_{M_q}H_{0,q}/4)^{1/3}$, where
$r_{M_q},H_{0,q}$ - values of $r_{M},H_0$, respectively, with due
regard for  {\bf qgcs}.
\\{\bf 3.1}. {\it The Stationary picture}.
\\From the start  of creation,  the primordial black hole, with
the mass $M$ and the event horizon area $A$, is considered in the
absence of absorption and radiation processes. It may be assumed
that such quantum  {\bf pbh} was generated immediately before the
onset of inflation, when there were no absorption and radiation
processes. On the other hand, {\bf 3.1} is completely consistent
with the paradigm in \cite{Calmet--1}, \cite{Calmet--2} presuming
the absence of Hawking radiation for quantum {\bf pbh}, with the
mass and the event horizon radius close to the Planckian values
$M\approx m_p,r_{M}\approx l_p$ .
\\As $\mu=const$ and {\bf pbh} we consider in the stationary state,
then, due to {\bf Remark  2.2} with regard for {\bf qgcs},
replacement $r_{M}\mapsto r_{M_q}$ in this formula leads to
replacement of $H_0\rightarrow H_{0,q}$, due to {\bf Remark  3.1.}
meeting the condition
\begin{equation}\label{SdS-1.mu.q}
\mu=(r_{M}H_0/4)^{1/3}=(r_{M_q}H_{0,q}/4)^{1/3}.
\end{equation}
Here $r_{M_q}=R_q$ from the general formula
(\ref{quant-grav.new}).
\\Based on the last formula and formulae
(\ref{quant-grav}),(\ref{QGC-1.Sch-old}),(\ref{quant-grav.new}) it
directly follows that
\begin{equation}\label{SdS-1.mu.q2}
H_{0,q}=H_0\exp \left(-\frac{1}{2}W\left( -%
\frac{1}{e}\left(\frac{M_{0}}{M}\right)^{2}\right) \right)=H_0\exp \left(-\frac{1}{2}W\left( -%
\frac{1}{e}\alpha_{r_{M}}\right) \right).
\end{equation}
Then {\bf qgc} for  the scale factor $a(\eta)$ (\ref{SdS-1.a})
\begin{eqnarray}\label{SdS-1.a.q}
a(\eta)\rightarrow a(\eta)_q\doteq -1/(H_{0,q}\eta)
=-1/(H_0\exp \left(-\frac{1}{2}W\left( -%
\frac{1}{e}\left(\frac{M_{0}}{M}\right)^{2}\right) \right)\eta)=\nonumber\\=-1/(H_0\exp \left(-\frac{1}{2}W\left( -%
\frac{1}{e}\alpha_{r_{M}}\right) \right)\eta)
=a(\eta)\exp\left(\frac{1}{2}W\left( -%
\frac{1}{e}\alpha_{r_{M}}\right) \right),\eta<0,
\end{eqnarray}
and for the Hubble parameter in general case $H(\eta)\doteq H$
\begin{eqnarray}\label{H.q0}
H=a'(\eta)/a^{2}(\eta)\mapsto H_q(\eta)=a'(\eta)_q/a^{2}(\eta)_q \
\end{eqnarray}
As directly follows from the last formula, in this pattern $H_0$
and $H$ are identically transformed, i.e. we have
\begin{eqnarray}\label{H.q}
H\mapsto
H_q=H\exp\left(-\frac{1}{2}W\left( -%
\frac{1}{e}\alpha_{r_{M}}\right) \right)
\end{eqnarray}
Because the potential energy of inflation $V$ is related to the
Hubble parameter $H$ by the Friedmann equation (formula (12.12) in
\cite{Rub-1}) $H^2=8\pi V/(3M_p^2)$,from (\ref{SdS-1.mu.q2}) we
can derive a ''shift'' for $V$ that is due to
quantum-gravitational corrections for the primordial Schwarzschild
black hole with the mass $M$ as follows:
\begin{eqnarray}\label{SdS-2.V}
[V=3M_p^2H^2/(8\pi)]\mapsto V_q=3M_p^2H_{q}^2/(8\pi)=\nonumber\\
=3\exp \left(-W\left( -%
\frac{1}{e}\alpha_{r_{M}}\right) \right)M_p^2H^2/(8\pi)=\exp \left(-W\left( -%
\frac{1}{e}\alpha_{r_{M}}\right) \right)V,
\end{eqnarray}
{\it where $V=V(\phi)$,$\phi$ is inflaton \cite{Rub-1}, and
the equation of motion (formula (12.5) in
\cite{Rub-1})
\begin{eqnarray}\label{Rub-new1}
\ddot{\phi}+3H\dot{\phi}+V'(\phi)=0
\end{eqnarray}
and Friedmann’s equation (formula (12.8) in \cite{Rub-1}) are
valid}:
\begin{eqnarray}\label{Rub-new2}
H^{2}=\frac{8\pi}{8m^{2}_p}(\frac{1}{2}\dot{\phi}^{2}+V(\phi)).
\end{eqnarray}
In a similar way, taking account of {\bf qgcs} for quantum {\bf
pbhs} (formulae (\ref{quant-grav},(\ref{quant-grav.new}), we can
find these ''shifts'' for all inflationary parameters, in
particular
\begin{eqnarray}\label{transform-1}
(a\sim H^{-1})\mapsto a_{q}\sim  H_{q}^{-1};
\nonumber\\
(V\sim H^2)\mapsto V_q\sim H^2_{q};
\nonumber\\
 (\dot{\phi}=-\frac{V'(\phi)}{3H}\sim H)\mapsto
\dot{\phi}_{q}=(-\frac{V'(\phi)}{3H})_{q}\sim H_{q};
\nonumber\\
(\dot{H}=\frac{1}{2m_{p}}(\frac{8\pi}{3V})^{1/2}V'(\phi)\dot{\phi}\sim
H^2)\mapsto\dot{H}_{q}=\frac{1}{2m_{p}}[(\frac{8\pi}{3V})^{1/2}V'(\phi)\dot{\phi}]_q\sim
H_{q}^{2};
\nonumber\\
(\ddot{\phi}=-\frac{m_{p}^{2}}{8\pi}(\frac{V''}{V}-\frac{1}{2}(\frac{V'}{V})^{2})H\dot{\phi}\sim
H^{2})\mapsto\ddot{\phi}_{q}\sim H_{q}^{2},...
\end{eqnarray}
 with retention of some part of them on the transformation
$H\mapsto H_q$,for example it is
 easy to check retention of the deceleration parameters
$\epsilon, \widetilde{\eta}$ \cite{Rub-1}:
\begin{eqnarray} \label{SdS-2.eps.q1}
\epsilon = -\frac{\dot{H}}{H^2}=m_p\frac{V'^{2}}{V^{2}}=\epsilon_q
= -\frac{\dot{H_{q}}}{H^2_{q}}=m_p\frac{V^{'2}_q}{V_q^{2}},
\nonumber\\
 \widetilde{\eta}=\frac{m_{p}^{2}}{8\pi}\frac{V''}{V}= \widetilde{\eta}_{q}.
\end{eqnarray}
Here in the last two formulae
(\ref{transform-1}),(\ref{SdS-2.eps.q1}) a point means
differentiation with respect to $t$, but a prime means
differentiation with respect to the field  $\phi$. Let us denote
the second deceleration parameter $\widetilde{\eta}$ instead of
$\eta$ \cite{Rub-1}, to avoid confusion with the conformal time.
\\ As we have $\epsilon=\epsilon_q,\widetilde{\eta}=\widetilde{\eta}_{q}$,the slow-roll conditions
(formula (\ref{SdS-2.eps.q1}) in \cite{Rub-1}) in the inflationary
scenario with regard to {\bf qgcs} for {\bf qpbhs} remains unaltered.
\\{\it Due to formula (\ref{quant-grav}), all the above-mentioned shifts of the inflationary parameters
generated by {\bf qgcs} for quantum  {\bf pbhs} in the pre-inflationary period may be series expanded in terms of
the small parameter $(M_0/M)^2$ (same $\alpha_{r_M}$)}.
\\
\\{\bf 3.2} {\it Black Hole Evaporation and {\bf qgcs}}
\\(This case studied in \cite{Shalyt-Proc} is given for completeness of the presentation.)
\\ Also,black holes are associated with the process of Hawking radiation (evaporation).
The primordial black holes are no exception.  In the general case
this process is considered only within the scope of a
semiclassical approximation (without consideration of the
quantum-gravitational effects). Because of this, it is assumed
that a primordial black hole may be completely evaporated
\cite{Frol}.
\\Still, in this pattern the situation is impossible due to the validity of {\bf GUP*} ({\bf
GUP}), (\ref{GUP.2}),(\ref{GUP})  and due to the formation of a
minimal (nonvanishing) Planckian remnant as a result of
evaporation (\ref{min}) \cite{GUPg2},\cite{Nou}.
\\We can compare the mass loss for a black hole
in this process when using a semiclassical approximation and with
due regard for {\bf qgcs}.
\\Let $M$ be the mass of a primordial black hole.
Then a loss of mass as a result of evaporation, according to the
general formulae, takes the following form (\cite{Frol},p.356):
\begin{equation}\label{Loss-1}
\frac{dM}{dt}\sim \sigma T^{4}_{\text{H}}A_{M},
\end{equation}
where $T_{\text{H}}$ - temperature of a black hole with the mass
$M$,$A_{M}$ - surface area of the event horizon of this hole
$A_{M}=4\pi r^{2}_{M}$, and
$\sigma=\pi^{2}k^{4}/(60\hbar^{3}c^{2})$ is the Stefan-Boltzmann
constant.
\\Using this formula for the same black hole but with regard to
{\bf qgcs}, we can get the mass loss $[dM/dt]_q$ in this case
\begin{equation}\label{Loss-1.q}
[\frac{dM}{dt}]_q\sim \sigma T^{4}_{\text{H},q}A_{M},
\end{equation}
where $T_{\text{H},q}$ - temperature of a black hole  with the
same mass $M$, when taking into consideration {\bf qgcs}
(\ref{quant-grav}).
\\For all the foregoing formulae associated with a random black hole having the mass $M$,
the following estimate is correct ((10.1.19) in \cite{Frol}):
\begin{equation} \label{Loss-2}
-\frac{dM}{dt}\sim b
(\frac{M_{p}}{M})^{2}(\frac{M_{p}}{t_p})^{2}N,
\end{equation}
where $b\approx 2.59\times 10^{-6}$, and $N$ is the number of the
states and species of particles that are radiated. The minus sign
in the left part of the last formula denotes that the mass of a
black hole diminishes as a result of evaporation, i.e. we have
$dM/dt<0$.
\\Unfortunately, the last formula is hardly constructive
as it is difficult to estimate the number $N$, especially at high energies
$E\simeq E_{p}$.
\\Nevertheless, using the terminology and symbols of this paper,
and also the results from \cite{Nou}, the formula (\ref{Loss-2})
for the mass loss by a black hole with regard to {\bf qgcs} may be
written in a more precise and constructive form. Really, according
to formula (45) in \cite{Nou}, within the scope of {\bf GUP*}
(\ref{GUP.2})  we will have
\begin{eqnarray} \label{Evapor-1}
\frac{dM}{dt}=-\frac{\gamma _{1}}{M^{2}l_{p}^{4}}\exp \left(-2W\left( -%
\frac{1}{e}\left( \frac{M_{0}}{M}\right) ^{2}\right)\right)\times\nonumber\\
\times\left( 1-\frac{%
8\gamma _{2}}{e\gamma _{1}}\left( \frac{M_{0}}{M}\right) ^{2}\exp
\left(-W\left( -\frac{1}{e}\left( \frac{M_{0}}{M}\right)
^{2}\right) \right) \right) ,
\end{eqnarray}
where $\gamma _{1}=\frac{\pi ^{2}}{480},$ $\gamma _{2}=\frac{\pi
^{2}}{16128}$.
\\The minus sign in the right side of the last formula means
the same as the minus sign in the left side of formula
(\ref{Loss-2}).
\\Due to (\ref{QGC-1.Sch-old}), formula (\ref{Evapor-1}) is of the following form:
\begin{eqnarray}\label{Evapor-2}
\frac{dM}{dt}=-\frac{\gamma _{1}}{M^{2}l_{p}^{4}}\exp(-2W\left( -%
\frac{1}{e}\alpha_{r_M}\right))
\times\nonumber\\
\times\left(1-\frac{8\gamma _{2}}{e\gamma _{1}}\alpha_{r_M}\exp
\left(-W\left(-\frac{1}{e}\alpha_{r_M}\right)\right)\right).
\end{eqnarray}
We can expand the right sides of formulae (\ref{Evapor-1}) and
(\ref{Evapor-2}) into a series in terms of the small parameter
$e^{-1}(M_{0}/M)^{2}=e^{-1}\alpha_{r_M}$ (formula (46) in
\cite{Nou}) that, proceeding from the deformation parameter
$\alpha_{r(M)}$, takes the form
\begin{eqnarray}\label{Evapor-3}
\frac{dM}{dt}=-\frac{\gamma
_{1}}{M^{2}l_{p}^{4}}\left(1+\frac{2}{e}\alpha_{r_M}+\frac{4}{e^{2}}\left( 1-\frac{2\gamma _{2}}{%
e\gamma _{1}}\right)\alpha_{r_M}^{2}+\frac{25}{3e^{3}}%
\left(1-\frac{72\gamma _{2}}{25e\gamma
_{1}}\right)\alpha_{r_M}^{3}+\ldots \right) .
\end{eqnarray}
Neglecting the last equation due to the time interval chosen,
e.g., due to $\triangle t=t_{infl}-t_{M}$,where $t_{infl}$--time
of the inflation onset and $t_{M}$-- time during which  the black
hole under study has been formed, formula  (\ref{PBH-Mass}), the
mass loss for a black hole with regard to {\bf qgcs} by the
inflation onset time may be given as
\begin{eqnarray}\label{Evapor-4.q}
\Delta_{Evap,q}M(t_{M},t_{infl})\doteq\int\limits_{t_{M}}^{t_{infl}}\frac{dM}{dt}=\nonumber\\
=-\int\limits_{t_{M}}^{t_{infl}}\frac{\gamma
_{1}}{M^{2}l_{p}^{4}}\left(1+\frac{2}{e}\alpha_{r_M}+\frac{4}{e^{2}}\left( 1-\frac{2\gamma _{2}}{%
e\gamma _{1}}\right)\alpha_{r_M}^{2}+\frac{25}{3e^{3}}%
\left(1-\frac{72\gamma _{2}}{25e\gamma
_{1}}\right)\alpha_{r_M}^{3}+\ldots \right) .
\end{eqnarray}
Next, we can determine the mass of a black hole after its
evaporation until the inflation onset with regard to {\bf qgcs}
\begin{eqnarray}\label{Evapor-4.2.q}
M_{Evap,q}(t_{M},t_{infl})\doteq
M+\Delta_{Evap,q}M(t_{M},t_{infl}).
\end{eqnarray}
In the pattern of a semiclassical approximation the
above-mentioned formulae are greatly simplified because in this
case $\alpha_{r_M}=0$ due to the absence of a minimal black hole.
\\Then in a semiclassical pattern formula (\ref{Evapor-4.2.q}),
with the use of the suggested formalism, takes the following form:
\begin{eqnarray}\label{Evapor-4.2}
M_{Evap}(t_{M},t_{infl})\doteq M+\Delta_{Evap}M(t_{M},t_{infl}),
\end{eqnarray}
where
\begin{eqnarray}\label{Evapor-4}
\Delta_{Evap}M(t_{M},t_{infl})=
\int\limits_{t_{M}}^{t_{infl}}\frac{dM}{dt}
=-\int\limits_{t_{M}}^{t_{infl}}\frac{\gamma
_{1}}{M^{2}l_{p}^{4}}.
\end{eqnarray}
Accordingly, for the radii $M_{Evap}(t_{M},t_{infl}),M_{Evap,q}$
we can get
\begin{eqnarray}\label{SdS-1.mu.evap.R}
r(M_{Evap})=2GM_{Evap}(t_{M},t_{infl}),
\nonumber\\
r(M_{Evap,q})=2G M_{Evap,q}(t_{M},t_{infl}).
\end{eqnarray}
In accordance with {\bf Remark 3.1}, we have
\begin{eqnarray}\label{SdS-1.mu.Evap}
\mu_{Evap}\doteq(r_{M_{Evap}}H_{0,Evap}/4)^{1/3}=(r_{M_{Evap},q}H_{0,Evap,q}/4)^{1/3};
\nonumber\\
H_{0,Evap,q}=\frac{r_{M_{Evap}}}{r_{M_{Evap},q}}H_{0,Evap}.
\end{eqnarray}
The right side of the last line in formula (\ref{SdS-1.mu.Evap})
gives the {\bf ''quantum-gravitational shifts''} (abbreviated as
{\bf qgs}) of the de Sitter Hubble parameter $H_{0}$ for
black holes evaporation process.
\\Substituting $H_{0,Evap,q}$ from (\ref{SdS-1.mu.Evap})
into formulae (\ref{SdS-2.V})--(\ref{SdS-2.eps.q1}) and so on, we
can obtain {\bf qgsc} for all cosmological parameters in the
inflationary scenario when a primordial black hole evaporates
before the inflation onset.
\\{\bf  Remark 3.2.}
\\By the present  approach we can consider the case of the
particle absorption by a {\bf pbh}. Let the
Schwarzschild-de~Sitter {\bf pbh} of the mass $M$ has the event
horizon area $A$.In \cite{Bek-1},\cite{Bek-2} ''a minimal
increment'' of the event horizon area for the black hole absorbing
a particle with the energy $E$ and with the size $R$:$\left(\Delta
A\right) _{\text{0}}\simeq 4l_p^{2}\left(\ln 2\right) ER$ has been
estimated within the scope of the Heisenberg Uncertainty Principle
{\bf HUP}. In quantum consideration we have $R\sim 2\delta X$ and
$E\sim \delta P$. Within the scope of GUP this '' minimal
increment'' is replaced by $\left(\Delta A\right)_{\text{0},q}$ as
follows (formula (27) in \cite{Nou}):
\begin{eqnarray}\label{QGC-1}
\left(\Delta A\right) _{\text{0},q}\approx 4l_p^{2}\ln 2\exp \left( -\frac{%
1}{2}W\left( -\frac{1}{e}\frac{A_{0}}{A}\right) \right)=
\nonumber\\
4l_p^{2}\ln 2\exp \left( -\frac{%
1}{2}W\left(-\frac{1}{e}\left( \frac{M_{0}}{M}\right)
^{2}\right)\right)=4l_p^{2}\ln
2\exp\left( -\frac{%
1}{2}W\left( -%
\frac{1}{e}\alpha_{r_M}\right)\right)
\end{eqnarray}
Assuming that an arbitrary increment of the event horizon area $A$
(same with the mass $M$) may be represented as a chain  of
''minimal increments'' (for quantum {\bf pbhs} with the mass close
to that of  the Planck's such an assumption is fairly justified)
for $\mu=const$ and any absorption we can compare in the given
approach the values of all cosmological parameters in the
semiclassical approximation and their ''shifts'' generated by {\bf
qgcs}.

\section{Quantum-Gravity Corrections for Appearance   Probabilities  PBHs in the Pre-Inflationary Era}

There is the problem of estimating the probability of occurrence
for {\bf pbh} with Schwarzschild-de~Sitter {\bf SdS} metric
(\ref{SdS-1}) in the pre-inflation epoch.
\\This problem has been studied in \cite{Prok} without due regard
for {\bf qgcs}. Let us demonstrate that consideration of {\bf
qgcs} in this case makes the probability of arising such  {\bf
pbhs} higher.
\\Similar to \cite{Prok}, it is assumed that in pre-inflation
period non-relativistic particles with the mass $m<M_p$ are
dominant (Section 3 in \cite{Prok}). For convenience, let us
denote the Schwarzschild radius $r_{M}$ by $R_S$.
\\When denoting, in analogy with \cite{Prok}, by $N(R,t)$
the number of particles in a \textit{comoving} ball with the
physical radius $R=R(t)$ and the volume $V_R$ at time $t$, in the
case under study this number (formula (3.9) in \cite{Prok}) will
have by {\bf qgc}:
\begin{eqnarray}\label{average:N}
N(R,t)\mapsto N(R,t)_q;\nonumber\\
 (\langle N(R,t)\rangle =
\frac{m_p^2H^2R^3}{2m})\mapsto (\langle N(R,t)_q\rangle =
\frac{m_p^2H^2_qR^3}{2m}).
\end{eqnarray}
Here the first part of the last formula agrees with formula (3.9)
in \cite{Prok}, whereas  $H,H_q$ in this case are in agreement
with formulae (\ref{H.q0}),(\ref{H.q}). And from (\ref{H.q}) it
follows that
\begin{equation}\label{SdS-1.mu.q2.prob}
\langle N(R,t)_q\rangle=\langle N(R,t)\rangle\exp\left(-W\left( -
\frac{1}{e}\left( \frac{M_{0}}{M}\right) ^{2}\right) \right).
\end{equation}
According to (\ref{quant-grav.new}), it is necessary to replace
the Schwarzschild radius $R_S$ by \\$R_{S,q}=R_S\exp
\left(\frac{1}{2}W\left(-\frac{1}{e}\left( \frac{M_{0}}{M}\right)
^{2}\right)\right)$.
\\Then from the general formula $N(R_S,t) = \langle N(R_S,t)\rangle + \delta
N(R_S,t)$, used because of the replacement of $R_S\mapsto
R_{S,q}$, we obtain an analog of (3.12) from \cite{Prok}
\begin{eqnarray}\label{deltaNcr,q}
\delta N >\delta N_{\rm cr,q}\doteq\frac{m_p^2R_{S,q}}{2m} -
\langle N(R_S,t)_q\rangle
          = \frac{m_p^2R_{S,q}}{2m}[1-(HR_S)^2]=\nonumber\\
=\frac{m_p^2R_{S}}{2m}[1-(HR_S)^2]\exp
\left(\frac{1}{2}W\left(-\frac{1}{e}\left( \frac{M_{0}}{M}\right)
^{2}\right)\right)=\delta N_{\rm cr}\exp
\left(\frac{1}{2}W\left(-\frac{1}{e}\left( \frac{M_{0}}{M}\right)
^{2}\right)\right).
\end{eqnarray}
In the last formula in square brackets we should have
$(H_qR_{S,q})^2$ instead of $(HR_S)^2$ but, as we consider the
case $\mu=const$, these quantities are coincident.
\\ It should be noted that here the following condition is used:
\begin{equation}\label{Hubble-R}
HR_S<1,
\end{equation}
i.e. Schwarzschild radius $R_S$ less than Hubble radius,
$R_S<R_H=1/H$.
\\ As we have $\exp
\left(\frac{1}{2}W\left(-\frac{1}{e}\left(\frac{M_{0}}{M}\right)
^{2}\right)\right)<1$, then
\begin{equation}\label{SdS-1.mu.q2.prob-new1}
\delta N_{\rm cr,q}<\delta N_{\rm cr}.
\end{equation}
Considering that for the formation of a Schwarzschild black hole
with the radius $R_S$ it is required that, due to statistical
fluctuations, the number of particles $N(R_S,t)$ with the mass $m$
within the black hole volume $V_{R_S}=4/3\pi R^{3}_S$ be in
agreement with the condition \cite{Prok}
\begin{equation}\label{SdS-1.mu.q2.prob-new2}
N(R_S,t)>R_SM_p^2/(2m),
\end{equation}
which, according to {\bf qgc} in the formula of
(\ref{quant-grav.new}), may be replaced by
\begin{equation}\label{SdS-1.mu.q2.prob-new2q}
N(R_{S,q},t)>R_{S,q}M_p^2/(2m)=\exp
\left(\frac{1}{2}W\left(-\frac{1}{e}\left( \frac{M_{0}}{M}\right)
^{2}\right)\right)R_S M_p^2/(2m).
\end{equation}
As follows from these expressions, with regard to {\bf qgc} for
the formation of {\bf pbh} in the pre-inflation period, the number
of the corresponding particles  may be lower than for a black hole
without such regard, leading to a higher probability of the
formation.
\\Such a conclusion may be made by comparison
of this probability in a semi-classical consideration (formula
(3.13) in \cite{Prok})
\begin{equation}\label{probability:BH formation}
P\big(\delta N(R_S,t) > \delta N_{\rm cr}(R_S,t)\big) =
\int_{\delta N_{\rm cr}}^\infty d(\delta N) P(\delta N)
\end{equation}
and with due regard for {\bf qgc}
\begin{equation}\label{probability:BH formation-q}
P\big(\delta N(R_{S,q},t) > \delta N_{\rm cr}(R_{S,q},t)\big) =
\int_{\delta N_{\rm cr,q}}^\infty d(\delta N) P(\delta N).
\end{equation}
Considering that in the last two integrals the integrands take
positive values and are the same, whereas the integration domain
in the second integral is wider due to
(\ref{SdS-1.mu.q2.prob-new1}), we have
\begin{eqnarray}\label{probability:BH formation-q2}
\int_{\delta N_{\rm cr,q}}^\infty d(\delta N) P(\delta N)=\nonumber\\
=\int_{\delta N_{\rm cr,q}}^{\delta N_{\rm cr}} d(\delta N)
P(\delta N)+\int_{\delta N_{\rm cr}}^\infty d(\delta N) P(\delta
N)>\int_{\delta N_{\rm cr}}^\infty d(\delta N) P(\delta N).
\end{eqnarray}
As follows from the last three formulae, in the case under study
the probability that the above-mentioned {\bf pbh} will be formed
is higher with due regard for {\bf qgc}.

\section{High Energy Deformations of Friedmann Equations}

Based on the obtained results, it is inferred that there is the
deformation (having a quantum-gravitational character) of the
Schwarzschild-de~Sitter metric and Friedmann Equations due to
these {\bf qgsc}. Indeed,substituting the expression $a(\eta)_q$
instead of $a$ into the Friedmann Equation ((2.4) in \cite{Rub-1})
without term with curvature
\begin{equation}\label{FE-1}
\frac{a'^{2}}{a^{4}}=\frac{8\pi}{3}G\rho,
\end{equation}
we can obtain the Quantum Deformation ({\bf QD}) \cite{Fadd} of
the Friedmann Equation due to {\bf qgcs} for {\bf pbh} in the
early Universe
\begin{equation}\label{FE-1.q}
\frac{a_q'^{2}}{a_q^{4}}=\frac{a'^{2}}{\exp\left(W\left( -%
\frac{1}{e}\alpha_{r_M}\right)\right)a^{4}} =\frac{8\pi}{3}G\rho
\end{equation}
or
\begin{eqnarray}\label{FE-1.q1}
\frac{a'^{2}}{a^{4}}
=\frac{8\pi}{3}G\rho\exp\left(W\left( -%
\frac{1}{e}\alpha_{r_M}\right)\right)\doteq\frac{8\pi}{3}G\rho_q,\nonumber\\
\rho_q\doteq\rho\exp\left(W\left( -%
\frac{1}{e}\alpha_{r_M}\right)\right)<\rho.
\end{eqnarray}
The last line in (\ref{FE-1.q1}) is associated with the fact that
the Lambert W-function $W(u)$ is negative for $u<0$.
\\Similarly, $(ij)$-components of the Einstein equations ((2.5) in
\cite{Rub-1})
\begin{equation}\label{FE-2}
2\frac{a''}{a^{3}}-\frac{a'^{2}}{a^{4}}=-\frac{8\pi}{3}Gp
\end{equation}
within the foregoing ({\bf QD}) are  replaced by
\begin{eqnarray}\label{FE-2.q}
2\frac{a_q''}{a_q^{3}}-\frac{a_q'^{2}}{a_q^{4}}=-\frac{8\pi}{3}Gp
\end{eqnarray}
or
\begin{eqnarray}\label{FE-2.q1}
2\frac{a''}{a^{3}}-\frac{a'^{2}}{a^{4}}=-\frac{8\pi}{3}Gp\exp\left(W\left( -%
\frac{1}{e}\alpha_{r_M}\right)\right)=-\frac{8\pi}{3}Gp_q,\nonumber\\
p_q\doteq p\exp\left(W\left( -%
\frac{1}{e}\alpha_{r_M}\right)\right)<p.
\end{eqnarray}
It should be noted that the equation of the covariant energy
conservation for the homogeneous background ((2.6) in
\cite{Rub-1})
\begin{eqnarray}\label{Covar}
\rho'=-3\frac{a'}{a}(\rho+p)
\end{eqnarray}
remains unaltered with replacement of $\rho\mapsto \rho_q,p\mapsto
p_q$.
\\So, in the pattern of {\bf 3.1} ({\it the stationary pattern}),
taking into consideration of {\bf qgcs} for {\bf pbhs} in the
pre-inflationary era decreases the initial values of the density
$\rho$ and of the pressure $p$ in Friedmann equations.
\\The above calculations are correct if, from the start,
we assume that a black hole (i.e., its event-horizon radius) is
invariable until the onset of inflation. But such a situation is
idealized because this period is usually associated with the
radiation and absorption processes
\\Then again for $\mu=const$ from formulae
(\ref{SdS-1.mu}),(\ref{SdS-1.a}) we have
\begin{eqnarray}\label{Shift-3.G}
H_{0,q}=H_{0}\frac{r_{M_{orig}}}{r_{M_{orig,q}}},
\nonumber\\
a(\eta)_q=a(\eta)\frac{r_{M_{orig,q}}}{r_{M_{orig}}}.
\end{eqnarray}
Substituting the expression $a(\eta)_q$ from formula
(\ref{Shift-3.G}) in all formulae (\ref{FE-1.q})--(\ref{Covar})
we obtain analogues of these formulae in the general case. In particular, for formula (\ref{FE-1.q}) we have
\begin{equation}\label{FE-1.q.G}
\frac{a_q'^{2}}{a_q^{4}}=
\frac{r^{2}_{M_{orig}}}{r^{2}_{M_{orig,q}}}\frac{a'^{2}}{a^{4}}
=\frac{8\pi}{3}G\rho
\end{equation}
Or, equivalently,
\begin{eqnarray}\label{FE-1.q1.G}
\frac{a'^{2}}{a^{4}}
=\frac{8\pi}{3}G\rho\frac{r^{2}_{M_{orig,q}}}{r^{2}_{M_{orig}}}=\frac{8\pi}{3}G\rho_q\nonumber\\
\rho_q\doteq\frac{r^{2}_{M_{orig,q}}}{r^{2}_{M_{orig}}}\rho.
\end{eqnarray}
In the same way as for formula (\ref{FE-2.q}), in this pattern for
the general quantum deformation $(ij)$-components of Einstein
equations by substitution of the value for $a(\eta)_q$   from the
formula  (\ref{Shift-3.G})  we obtain
\begin{eqnarray}\label{FE-2.q.G}
2\frac{a_q''}{a_q^{3}}-\frac{a_q'^{2}}{a_q^{4}}=-\frac{8\pi}{3}Gp
\end{eqnarray}
or
\begin{eqnarray}\label{FE-2.q1.G}
2\frac{a''}{a^{3}}-\frac{a'^{2}}{a^{4}}=-\frac{8\pi}{3}Gp\frac{r^{2}_{M_{orig,q}}}{r^{2}_{M_{orig}}}
=-\frac{8\pi}{3}Gp_q,\nonumber\\
p_q\doteq \frac{r^{2}_{M_{orig,q}}}{r^{2}_{M_{orig}}}p.
\end{eqnarray}
It is clear that, in this most general pattern, the covariant energy
conservation for the homogeneous background ((2.6) in
\cite{Rub-1})
\begin{eqnarray}\label{Covar.G}
\rho'=-3\frac{a'}{a}(\rho+p)
\end{eqnarray}
remains unaltered with replacement of  $\rho\mapsto
\rho_q,p\mapsto p_q$.

\section{The Quantum Fluctuations and Cosmological Perturbations
Corrections Generated by {\bf qgcs} for {\bf qpbhs}.
Non-Gaussianity Enhancement. The Onset}

\subsection{General Remarks}

It is known that inflationary cosmology is characterized by {\it
cosmological perturbations} of different nature (scalar, vector,
tensor) \cite{Rub-1},\cite{Wein-Cosm},\cite{Mukhanov-Cosm}, though
vector perturbations are usually ignored as they die out fast.
\\It is clear that, as {\bf qgcs} for {\bf pbhs} in the early Universe
cause shifts of the inflationary parameters, they inevitably lead
to {\bf corrections} of the cosmological perturbations on
inflation.
\\ The following Remark, with the inferences used in
the previous section, holds true:
\\
\\{\bf Remark  6.1}
{\it In what follows the calculations of the above-corrections are
performed using formulae
(\ref{SdS-1.mu.q2}),(\ref{SdS-1.a.q},(\ref{transform-1})}.
\\Further, without loss of generality, it is assumed that {\bf qpbhs} are associated with {\it the Stationary pattern} {\bf 3.1}
in supposition that the processes of black hole evaporation and
particle absorption have been finished by the moment of the
inflation onset.

\subsection{Commentary on Corrections for  Quantum Fluctuations and  Cosmological  Perturbations}

Specifically, in the case of scalar cosmological perturbations
consideration of the indicated {\bf qgcs} for the rest of the
Einstein equations (formulae (2.74)--(2.76) in \cite{Rub-1}) in
case {\bf 3.1}  in virtue of {\bf Remark 6.1} gives
\begin{eqnarray}\label{Scal-pert-1.qgcs}
\Delta\Phi-3\frac{a'}{a}\Phi'-3\frac{a'^2}{a^2}\Phi=4\pi Ga^2\exp\left(W\left( -%
\frac{1}{e}\alpha_{r_M}\right) \right)\cdot\delta\rho_{tot};
\nonumber\\
\Phi'+\frac{a'}{a}\Phi=-4\pi Ga^2\exp\left(W\left( -%
\frac{1}{e}\alpha_{r_M}\right)
\right)\cdot[(\rho+p)\upsilon]_{tot};
\nonumber\\
\Phi''+3\frac{a'}{a}\Phi'+(2\frac{a''}{a}-\frac{a'^2}{a^2})\Phi=4\pi
Ga^2\exp\left(W\left( -\frac{1}{e}\alpha_{r_M}\right)
\right)\cdot\delta p_{tot},
\end{eqnarray}
{\it where in conformal Newtonian gauge the metric with scalar
perturbations (formula (2.69) in \cite{Rub-1})is  given as
\begin{eqnarray}\label{Perturb-metric}
ds^{2}=a^{2}(\eta)[(1+2\Phi)d\eta^{2}-(1+2\Psi)d{\bf x}^{2}]
\end{eqnarray}
and $\Phi$ is the Newtonian potential, $\Psi$ is the field
in terms of which in this gauge the spatial components $h_{ij}$ of scalar perturbations are expressed:
$h_{ij}=-2\Psi\delta_{ij}$ (formula (2.67) in \cite{Rub-1})}.
\\Here in the right sides of all lines in the last formula the
scale factor $a$ is taken with regard to {\bf qgcs} from formula
(\ref{SdS-1.a.q}), i.e., $a=a(\eta)_q$. In the left sides of these
lines additional factors of the type
\\$\exp\left(\frac{1}{2}W\left( -%
\frac{1}{e}\alpha_{r_M}\right) \right),\exp\left(W\left( -%
\frac{1}{e}\alpha_{r_M}\right) \right),...$ are cancelled out
because they are independent of $\eta$. This is so in the general
case when taking in consideration {\bf qgcs} for the {\bf pbhs}
formed in the pre-inflationary era (for all types of the
cosmological perturbations, not only for those of the scalar
type).
\\According to this remark, under the linearized form
of the gauge transformations (formulae (2.31) in \cite{Rub-1}), spatial
components of the metric perturbation transform are retained due
to inclusion of {\bf qgcs} (\cite{Rub-1},p.30):
\begin{eqnarray}\label{Scal-pert-2.qgcs}
\widetilde{h}_{ij}=h_{ij}-2\partial_{i}\partial_{j}\sigma-\frac{a'}{a}\delta_{ij}\sigma'.
\end{eqnarray}
And {\bf qgcs} deform correspondingly the metric with scalar
perturbations in the conformal Newtonian gauge (formulae (2.69) in
\cite{Rub-1}):
\begin{eqnarray}\label{Scal-pert-3.qgcs}
\{ds^{2}=a^2(\eta)[(1+2\Phi)d\eta^2-(1+2\Psi)d\mathbf{x}^2]\}\mapsto
a^2(\eta)_q[(1+2\Phi)d\eta^2-(1+2\Psi)d\mathbf{x}^2]=
\nonumber\\
=a^2(\eta)\exp\left(W\left( -%
\frac{1}{e}\alpha_{r_M}\right)
\right[(1+2\Phi)d\eta^2-(1+2\Psi)d\mathbf{x}^2].
\end{eqnarray}
At the same time, the well-known Mukhanov–Sasaki equation
\cite{Mukh-new1},\cite{Sasaki-1},\cite{Liddle-1}
\begin{eqnarray}\label{Mukh-Sas}
\frac{1}{z}\frac{d^2z}{d\eta^2}=2a^{2}H^{2}(1+\epsilon-\frac{3}{2}\widetilde{\eta}
+\frac{1}{2}\widetilde{\eta}^{2}-\frac{1}{2}\epsilon\widetilde{\eta}+\frac{1}{2H}\frac{d\epsilon}{dt}-\frac{1}{2H}\frac{d\widetilde{\eta}}{dt}),
\end{eqnarray}
holds for these {\bf qgcs}.
\\This is inferred from the fact that formula (\ref{transform-1}) may be completed with two lines
\begin{eqnarray}\label{transform-new-1}
\dot{\epsilon}\sim H;
\nonumber\\
\dot{\widetilde{\eta}}\sim H.
\end{eqnarray}
As usual, in (\ref{Mukh-Sas})  $z\equiv a\dot{\phi}/H$
\cite{Liddle-1}.
\\Then, for convenience to emphasize the use of
quantum {\bf pbh} with the mass $M$ from the start, we introduce
the following designations:
\begin{eqnarray}\label{SdS-1.a.q-NEW}
a(\eta)_{M}\doteq a(\eta)_q,H_{M}\doteq H_{q};V_M\doteq V_q,...
\end{eqnarray}
Let us consider, as in \cite{Bartolo-1}, quantum fluctuations of a
generic scalar field during the de Sitter stage (p.121 in
\cite{Bartolo-1}). Similar to section 2.3 of \cite{Bartolo-1}, we
represent the scalar field
 $\phi_0$ in the form $\chi(\tau, \bf x)$ as follows:
\begin{eqnarray}
\chi(\tau, \bf x)=\chi(\tau)+\delta \chi(\tau, \bf x)\, ,
\end{eqnarray}
where $\chi(\tau)$ is the homogeneous classical value of the
scalar field, $\delta \chi$ are its fluctuations and, in line with
\cite{Bartolo-1},  $\tau$  is the the conformal time  instead of
$\eta$.
\\Then, specifically for quantum fluctuations of a generic scalar
field during the de Sitter stage (section 2.3 in \cite{Bartolo-1}),
the Klein--Gordon equation, which   gives in an unperturbed FRW
Universe (formula (40) in \cite{Bartolo-1})
\begin{equation}
\chi^{\prime\prime}+2\mathcal{H}\chi^\prime = - a^2\frac{\partial
V}{\partial\chi},
\end{equation}
due to {\bf Remark 6.1},  remains unaltered with deformation
(\ref{SdS-1.mu.q2})--(\ref{transform-1}),(\ref{SdS-1.a.q-NEW})
\begin{equation}
\{\chi^{\prime\prime}+2\mathcal{H}_M\chi^\prime = -
a_M^2\frac{\partial V_M}{\partial
\chi}\}\equiv\{\chi^{\prime\prime}+2\mathcal{H}\chi^\prime = -
a^2\frac{\partial V}{\partial\chi}\}.
\end{equation}
Here in the usual way we have
$\mathcal{H}=a^{\prime}/a=a_M^{\prime}/a_M=\mathcal{H}_M$.
\\But the above-mentioned deformation generated by
{\bf qgcs} for quantum {\bf pbhs} not always retains the physical
quantities  arising in this consideration.
\\Because in this case such an important quantity as
the power-spectrum acquires  multiplicative increments. In
particular,  in the case of the de Sitter phase and of a very
light scalar field $\chi$, with $m_\chi \ll 3/2 H$ (formula (64)
in \cite{Bartolo-1}), the power-spectrum on superhorizon scales in
this pattern is changed
\begin{eqnarray}\label{PS-1}
\{\mathcal P_{\delta \chi}(k)=\left( \frac{H}{2 \pi} \right)^2
\left(\frac{k}{aH}\right)^{3-2\nu_\chi}\}\Rightarrow \{\mathcal
P_{\delta\chi,M}(k)=\left(\frac{H_{M}}{2 \pi} \right)^2 \left(
\frac{k}{a_MH_{M}}\right)^{3-2\nu_\chi}\}=\nonumber\\
=\exp\left(-W\left( -%
\frac{1}{e}\left( \frac{M_{0}}{M}\right)^{2}\right)\right)\mathcal
P_{\delta \chi}(k).
\end{eqnarray}
In the last formula we use the obvious equality to the condition
$aH=a_MH_{M}$  and the fact that the quantity $\nu_\chi$ satisfies
the condition (formula (57) in \cite{Bartolo-1})
\begin{eqnarray}\label{nu}
\frac{3}{2}-\nu_\chi\simeq \eta_\chi=(m_\chi^2/3H^2).
\end{eqnarray}
As for the effective mass $m_\chi$ of the scalar field $\chi$ we
have the equality $m_\chi^2=\partial^2V/\chi^2$, from formula
(\ref{SdS-2.V}) it follows that the exponent $3-2\nu_\chi$ in
formula (\ref{PS-1})  is also retained with this deformation.
Since the quantity  $\exp\left(-W\left( -%
\frac{1}{e}\left( \frac{M_{0}}{M}\right)^{2}\right)\right)>1$,
with due regard for {\bf qgcs} in the de Sitter stage, magnitude of the power-spectrum is growing.
\\ Similar results are true also for the case of
 quantum fluctuations of a generic scalar field in the quasi–de Sitter
 stage (Section 2.4 in \cite{Bartolo-1}). We obtain an analog of the shift in formula
 (\ref{PS-1})
\begin{eqnarray}\label{PS-1.New}
\{\mathcal P_{\delta \chi}(k)\simeq \left( \frac{H}{2 \pi}
\right)^2 \left(\frac{k}{aH}\right)^{3-2\nu_\chi}\}\Rightarrow
\{\mathcal P_{\delta\chi,M}(k)\simeq\left(\frac{H_{M}}{2 \pi}
\right)^2 \left(
\frac{k}{a_MH_{M}}\right)^{3-2\nu_\chi}\simeq\nonumber\\
\simeq\exp\left(-W\left( -%
\frac{1}{e}\left( \frac{M_{0}}{M}\right)^{2}\right)\right)\mathcal
P_{\delta \chi}(k)\}.
\end{eqnarray}
Though, as distinct from \ref{nu}), formula (72) in
\cite{Bartolo-1} is true
\begin{eqnarray}\label{nu.1}
\frac{3}{2}-\nu_\chi\simeq\eta_\chi-\epsilon,
\end{eqnarray}
where $\epsilon$ is the deceleration parameter  from formula
(\ref{SdS-2.eps.q1}) characterized by $\epsilon=\epsilon_M\doteq
\epsilon_q$.From this it follows that formula  (\ref{PS-1.New})
is valid, whereas the power-spectrum for quantum
fluctuations of a generic scalar field in the quasi–de Sitter stage
is growing similar to the de Sitter stage.
\\
\\{\bf Remark  6.2}
\\ From {\bf Remark  2.3} it directly follows that right sides of formulae
(\ref{PS-1}),(\ref{PS-1.New})are series expanded  in terms of the small parameter
$\alpha_{r_M}$:
\begin{eqnarray}\label{PS-1.row}
\mathcal P_{\delta\chi,M}(k)
\simeq\exp\left(-W\left( -%
\frac{1}{e}\alpha_{r_M}\right)\right)\mathcal P_{\delta \chi}(k)=
\nonumber\\
=(1+\mathbf{\omega_1}\alpha_{r_M}+\mathbf{\omega_2}\alpha^{2}_{r_M}+\mathbf{\omega_3}\alpha^{3}_{r_M}+...)P_{\delta
\chi}(k),
\end{eqnarray}
where the first (leading) term in this series corresponds to the initial value of this power-spectrum (without the
indicated {\bf qgcs}), whereas the other terms  -- to the corrections of different orders  for these {\bf qgcs},
which are the greater the lower the size of the initial {\bf pbh}, i.e. the lower the mass  $M$ (or same the radius $r_M$).
\\ From formula  (\ref{quant-grav}) we can easily derive explicit values of the coefficients
 $\omega_i,i\geq 1$. Indeed, because
\begin{eqnarray}\label{PS-1.row2}
(1+\mathbf{\omega_1}\alpha_{r_M}+\mathbf{\omega_2}\alpha^{2}_{r_M}+\mathbf{\omega_3}\alpha^{3}_{r_M}+...)=\left(
1+\frac{1}{2e}\alpha_{r_M}+\frac{5}{8e^{2}}\alpha_{r_M}
^{2}+\frac{49}{48e^{3}}\alpha_{r_M}^{3}+\ldots \right)^2,
\end{eqnarray}
we have
$\mathbf{\omega_1}=1/e,\mathbf{\omega_2}=3/2e^{2},...$.
\\ In the right side of  (\ref{PS-1.row2}) we have the squared series from the right side of formula
(\ref{quant-grav})  in terms of the small parameter $\alpha_{r_M}$.
\\ It is clear that  we can also express as a series  in terms of a dimensionless small parameter
$\alpha_{r_M}$  the right sides in other formulae too, specifically in
(\ref{FE-1.q})--(\ref{FE-2.q1}),(\ref{Scal-pert-1.qgcs}),(\ref{Scal-pert-3.qgcs}), etc.
In every case, a part of this series going after the leading term detects the  deviation from  the
semi-classical approximation  mode in the approach to quantum gravity within the scope   of GUP (\ref{GUP}).

\subsection{Enhancement of Non-Gaussianity}

It is understandable, when  non-Gaussianities are involved in the
process of inflation, the above-mentioned {\bf qgcs} for quantum
{\bf pbhs} should change the parameters of these
non-Gaussianities. It is required to know what are the changes.
Deviation from the Gaussian distribution for the random field
$g(\mathbf{x})$ is a nonzero value of the three-point correlator
((6.30) in \cite{Liddle-1})
\begin{eqnarray}\label{Bispectrum}
\langle g_{\mathbf{k_1}}g_{\mathbf{k_2}}g_{\mathbf{k_3}} \rangle =
(2\pi)^{3}\delta_{\mathbf{k_1}+\mathbf{k_2}+\mathbf{k_3}}^{3}\textit{B}_{g}(k_1,k_2,k_3),
\end{eqnarray}
where $g_{\mathbf{k_i}},i=1,2,3$ represent the Fourier transform
$g(\mathbf{x})$ in the momentum $k_i$ and the quantity $\textit{B}_{g}$,
named the {\it bispectrum}, is a measure of non-Gaussinity for the random field $g(\mathbf{x})$.
\\ We can show that, with due regard for the foregoing  {\bf qgcs}, the absolute value of the {\it bispectrum} $\textit{B}_{g}$
in the inflation pattern for different random fields is growing.
\\ Let us consider several examples:
\\
\\{\bf 6.3.1 Non-Gaussianity {\bf qgcs}-Correction of Field Perturbations in Different Patterns}
\\
\\{\bf Non-Gaussianity correction from the self-interaction of the
field}
\\
\\In this case the  non-Gaussianity arises
from the self-interaction of the field (\cite{Lyth} and {\bf
24.4.2} in \cite{Liddle-1}). Neglecting the metric
perturbation at the initial stage and considering the effect of keeping the quadratic
term in the perturbed field equation, we can obtain (formula (24.39) in \cite{Liddle-1}) the following:
\begin{eqnarray}\label{PFE}
\ddot{\delta}\phi+3H_{\ast}\dot{\delta}\phi-a^{-2}\nabla^{2}\delta\phi+\frac{1}{2}V^{'''}_{\ast}(\delta\phi)^{2}=0.
\end{eqnarray}
Here the asterisk denotes a value of the corresponding quantity during
inflation which is taken to be constant, and the dot,  as usual,
denotes $\partial/\partial t$.
\\ Then the quantity
$\textit{B}_{g}$  in this case (formula (24.43) in \cite{Liddle-1}) takes the form
\begin{eqnarray}\label{Bispectrum-2}
\textit{B}^{self}_{\delta\phi}\sim H^{2}_{\ast}V^{'''}_{\ast}.
\end{eqnarray}
However, considering the  {\bf qgcs} under study, due to formulae
(\ref{H.q}),(\ref{transform-1}) and
(\ref{QGC-1.Sch}),(\ref{QGC-1.Sch-old}), the quantities in the right side of the last formula
 are transformed as follows:
\begin{eqnarray}\label{Bispectrum-2.1}
H^{2}_{\ast}\rightarrow H^{2}_{\ast,M}=\exp \left(-W\left( -%
\frac{1}{e}\alpha_{r_M}\right) \right)H^{2}_{\ast};
\nonumber\\
V^{'''}_{\ast}\rightarrow V^{'''}_{\ast,M}=\exp \left(-W\left( -%
\frac{1}{e}\alpha_{r_M}\right) \right))V^{'''}_{\ast}.
\end{eqnarray}
So, considering these {\bf qgcs}, we find that the {\it
bispectrum} $\textit{B}^{self}_{\delta\phi,M}$ in this case is
\begin{eqnarray}\label{Bispectrum-3}
\textit{B}^{self}_{\delta\phi,M}=\exp \left(-2W\left( -%
\frac{1}{e}\alpha_{r_M}\right)
\right)\textit{B}^{self}_{\delta\phi}.
\end{eqnarray}
As $\exp \left(-W\left( -%
\frac{1}{e}\alpha_{r_M}\right) \right)>1$,from the last relation it follows that
\begin{eqnarray}\label{Bispectrum-3.1}
|\textit{B}^{self}_{\delta\phi,M}|>|\textit{B}^{self}_{\delta\phi}|.
\end{eqnarray}
This means that with  regard to {\bf qgcs}  for quantum {\bf pbhs}
non-Gaussianity is enhanced.
\\ According to  {\bf Remark  6.2}, $\exp \left(-2W\left( -%
\frac{1}{e}\alpha_{r_M}\right) \right)$ can also be series expanded in terms of the dimensionless small
parameter $\alpha_{r_M}$ by squaring of the left side in the formula (\ref{PS-1.row2}) and by calculating the coefficients
for the corresponding powers $\alpha_{r_M}$. As noted in  {\bf Remark  6.2}, nontrivial terms of this series
generate {\bf qgcs} to the semiclassical {\it bispectrum}
$\textit{B}^{self}_{\delta\phi}$.
\\
\\{\bf The Non-Gaussianity Correction of Field Perturbations from  Gravitational interactions}
\\
\\ In the above analysis of non-Gaussianity there was no
metric perturbation because the gravitational interaction
was''excluded''. When it is taken into consideration, the non-Gaussianity in this case arises from the metric perturbation
and the {\it bispectrum} takes the following form (\cite{Seery} and formula (24.46) in \cite{Liddle-1}):
\begin{eqnarray}\label{Bispectrum-grav}
\textit{B}^{grav}_{\delta\phi}\sim -\frac{1}{8}
H^{4}_{\ast}(\frac{V'}V).
\end{eqnarray}
With regard to {\bf qgcs}, from formula (\ref{H.q}) we can derive
\begin{eqnarray}\label{Bispectrum-grav,M}
\textit{B}^{grav}_{\delta\phi}\rightarrow
\textit{B}^{grav}_{\delta\phi,M}=\exp\left(-2W\left( -%
\frac{1}{e}\left(\frac{M_{0}}{M}\right)^{2}\right)\right)
\textit{B}^{grav}_{\delta\phi}.
\end{eqnarray}
Note that (\ref{Bispectrum-grav,M}) is valid only in the case when {\bf qgcs}-correction of the metric is not taken
into consideration  or such correction is considered to be vanishingly small.  In this case from formulae
(\ref{Bispectrum-3}),(\ref{Bispectrum-grav,M}) we can obtain
\begin{eqnarray}\label{Bispectrum-new}
\frac{\textit{B}^{self}_{\delta\phi,M}}{\textit{B}^{self}_{\delta\phi}}=\frac{\textit{B}^{grav}_{\delta\phi,M}}{\textit{B}^{grav}_{\delta\phi}}
=\exp\left(-2W\left( -%
\frac{1}{e}\alpha_{r_M}\right)\right).
\end{eqnarray}
Based on the last formula,  it can be concluded that
\\{\it growth of non-Gaussianity of the field perturbations in
inflation generated by {\bf qgcs} for a  primordial black hole in
the pre-inflation era is independent of the pattern (with or without regard for the metric perturbation) considered if
{\bf qgcs}-correction of the metric may be neglected.}
\\
\\{\bf 6.3.2 Non-Gaussianity {\bf qgcs}-Correction for  the Tensor Primordial Perturbations}
\\
\\ As seen, the right side of   (\ref{Bispectrum-new}) arises with regard to the
{\bf qgcs}-correction  as a factor of the enhanced non-Gaussianity and the tensor primordial
perturbations.
\\ Actually, denoting in this case  {\it bispectrum} as $\textit{B}_{h^{TT}_{ij}}$, where by formula (5.43) in
\cite{Rub-1} we have
\begin{eqnarray}\label{Rub-tens}
h^{TT}_{ij}(\eta,\mathbf{k})=\sum_{\substack{A=+,\times}}e^{(A)}_{ij}h^{(A)}(\eta,\mathbf{k}),
\end{eqnarray}
in the left side the symmetric transverse traceless tensor
$h^{TT}_{ij}$ ((2.42) in \cite{Rub-1}) has helicity 2, whereas the right side represents expansion of this tensor
by the sum over polarization.
\\ According to formulae (1.1) in \cite{Maldacena} and  (24.63) in \cite{Liddle-1}, we have
\begin{eqnarray}\label{Bispectrum-tens}
\textit{B}_{h^{TT}_{ij}}\sim \frac{H^{4}_{\ast}}{m^4_p}.
\end{eqnarray}
Whence, in analogy with formula (\ref{Bispectrum-new}), it directly follows that
\begin{eqnarray}\label{Bispectrum-tens.M}
\frac{\textit{B}_{h^{TT}_{ij},M}}{\textit{B}_{h^{TT}_{ij}}}
=\exp\left(-2W\left( -%
\frac{1}{e}\alpha_{r_M}\right)\right).
\end{eqnarray}
 From the start In this section it has been assumed that this quantum
{\bf pbh} in the pre-inflationary era is considered in the state
{\bf 3.1}. {\it the stationary pattern}. Still, we can obtain similar results for other processes proceeding before the onset
of inflation, specifically for  {\bf 3.3} {\it black hole evaporation}.

\section{Conclusion and  Further Steps}

In this way it has been demonstrated that, within the scope of natural assumptions, the
{\bf qgcs} calculated for  {\bf pbhs} arising  in the pre-inflationary epoch contribute significantly to the
inflation parameters, enhancing non-Gaussianity in the case of cosmological perturbations.
Besides, withy due regard for these {\bf qgcs}, the probability of  arising {\bf pbhs} is higher.
\\Based on the results of this paper,
the following steps may be planned to study the corrections of
cosmological parameters and cosmological perturbations due to {\bf
qgcs} for {\bf pbhs} in the pre-inflationary era:
\\
\\{\bf 7.1} Comparison of the results obtained
in Section 6 with the experimental data accumulated by space
observatories: (Planck Collaboration), (WMAP Collaboration)
\cite{Planck-1},\cite{WMAP-1},\cite{WMAP-2}.
\\
\\{\bf 7.2} Elucidation of the fact, how closely the author's results are related to general approaches to  inclusion of
the quantum-gravitational effects in studies of inflationary perturbations (for example,
\cite{Kiefer-1}--\cite{Brus});
\\
\\{\bf 7.3} Elucidation of the possibility to extend the obtained results to other types of
 quantum {\bf pbhs}, in particular  to {\bf pbh}  with the mass $M$, with the electric charge $Q$ but without
rotation in the Reissner-Nordstr\"om (RN) Metric (for the normalization
$c=\hbar=G=1$) \cite{Frol},\cite{Prof-1}
\begin{equation}\label{RN}
ds^2 = \left(1 + \frac{2M}{r} + \frac{Q^2}{r^2}\right)dt^2 -
\left(1 - \frac{2M}{r} + \frac{Q^2}{r^2}\right)^{-1}dr^2- r^2
d^2\Omega.
\end{equation}
IN this case the event horizon radius $r_{\pm,RN}$ and the temperature
$T_{RN}$ of such a hole  respectively take the following forms:
\cite{Frol},\cite{Prof-1}
\begin{equation}\label{R}
r_{\pm,RN}= M \pm \sqrt{M^2-Q^2}.
\end{equation}
and
\begin{equation}\label{eq:rn-temperature}
T_{RN}=\frac{\left(M^2-Q^2\right)^{1/2}}{2\pi\left(M+\left(M^2-Q^2\right)^{1/2}\right)^2}.
\end{equation}
Selecting in formula (\ref{R}) for the radius $r_{RN}$ of such {\bf pbh}
the value $r_{RN}=r_{+}$  and assuming, similar to \cite{Page-1}, that $Q/M\ll 1$,
we can obtain that  the (RN) metric (\ref{RN}) for small $M$ represents, to a high accuracy, the Schwarzschild metric
(\ref{BH-0}) and the Schwarzschild-de~Sitter metric (\ref{SdS-1})
with the corresponding formulae for the  event horizon radius $r_{RN}$
and for the temperature $T_{RN}$, which are close to the corresponding Schwarzschid's  from Section 2.
In this way for  the
quantum PBH with (RN) metric derivation of the results similar to those given in this work (inclusion of
{\bf qgcs}) is relevant, at least for the case $Q/M\ll 1$.

\begin{center} {\bf Conflict of Interests}
\end{center}
The author declares that there is no conflict of interests
regarding the publication of this work.

\end{document}